\documentstyle[12pt,twoside,fleqn,espcrc1]{article}

\newcommand{\AmS}{{\protect\the\textfont2
  A\kern-.1667em\lower.5ex\hbox{M}\kern-.125emS}}

\title{Coherent electromagnetic heavy ion reactions: (1) exact treatment of
pair production and ionization; (2) mutual Coulomb dissociation}

\author{A. J. Baltz\address{Physics Department, 
        Brookhaven National Laboratory, \\ 
        Upton, New York 11973, USA} }

\begin{document}
\maketitle

\begin{abstract}
Some recent theoretical results on coherent electromagnetic processes in
ultrarelativistic heavy ion reactions are surveyed.
\end{abstract}

\section{INTRODUCTION: LARGE ELECTROMAGNETIC CROSS SECTIONS}

In ultrarelativistic heavy ion collisions, Coulomb induced cross sections are
huge, much larger than geometric.  For the RHIC case of 100 GeV $\times$
100 GeV colliding gold ions the predicted cross section for bound-electron
positron pairs is about about 110 barns\cite{brw2}.  The corresponding cross
section for continuum electron-positron pairs has recently been recalculated
to be 34,000 barns\cite{he}, consistent with the result of the classic formula
of Landau and Lifshitz\cite{ll}.  The cross section for Coulomb
dissociation of the nucleus is about 95 barns\cite{brw1}, and the cross section
for ionization of a single electron on one of the ions is about 100,000
barns\cite{ajbx}.

\section{THE $\delta(z-t)$ POTENTIAL AND ITS SOLUTION}
If one works in the appropriate gauge \cite{brw}, then
the Coulomb potential produced by an ultrarelativistic particle (such as a
heavy ion) in uniform motion can be expressed in the following form\cite{ajb}
\begin{equation}
V(\mbox{\boldmath $ \rho$},z,t)
=-\delta(z - t) \alpha Z_P(1-\alpha_z)
\ln{({\bf b}-\mbox{\boldmath $ \rho$})^2 \over b^2 }.
\end{equation}
${\bf b}$ is the impact parameter, $\alpha_z$ is the Dirac matrix,
and the other quantites are the usual coordinates and charge of the moving ion.
This the physically relevant ultrarelativistic potential since it was 
obtained by ignoring terms in
$({\bf b} - \mbox{\boldmath $\rho$}) / \gamma^2$\cite{ajb}\cite{brw}.  

It was shown in Ref.\cite{ajbl} that the $\delta$ function allows the
Dirac equation to be solved exactly at the point of interaction, $z = t$.
Exact amplitudes then take 
the same form as perturbation theory amplitudes, but with
an effective potential to represent all the higher order effects exactly
\begin{equation}
V(\mbox{\boldmath $ \rho$},z,t)
=-i \delta(z - t) (1-\alpha_z)
( e^{-i \alpha Z_P \ln{({\bf b}- \mbox{\boldmath $ \rho$})^2 }} - 1 )
\end{equation}
in place of the potential of Eq.(1).  

\subsection{Bound electron positron pair production}
Early nonperturbative coupled channels calcutions at $\gamma = 2.3$ found
enhancements of some two orders of magnitude over corresponding perturbation
theory calculations of Pb + Pb at small but non-hadronic impact
parameters\cite{rum}.  Improved coupled channels calculations at RHIC energies
$(\gamma = 23,000)$ indicated approximately a 10\% enhancement over
perturbation theory results\cite{brw3}.  This result was superseded by
calculations employing the exact delta function solution, Eq.(2), resulting
in a 2--3\% reduction from the perturbation theory result\cite{ajbl}

\subsection{Ionization}
From impact dependent probabilities for ionization also computed with Eq.(2),
cross sections have been calculated\cite{ajbx} for various ion-ion collision
combinations in the form $\sigma =  A \ln \gamma + B$
where $A$ and $B$ are constants for a given ion-ion pair and
$\gamma\ (=1/\sqrt{1-v^2})$ is the
relativistic factor one of the ions seen from the rest frame of the
other.

The agreement with the Anholt and Becker
calculations\cite{ab} in the literature is good for the lighter species for
both $A$ and $B$.  For heavier ion collisions such as Pb + Pb it is
the perturbative energy dependent term $A$ that shows the most discrepancy
with Anholt and Becker being about 60\% higher.  Perhaps this discrepancy
is due to the fact that Anholt and Becker use approximate relativistic bound
state wave functions and the present calculation utilizes exact
Dirac wave functions for the Pb bound states. 

CERN SPS data of Pb with a single electron
impinging on a Au target has recently been
published by Krause et al.\cite{kr}.  
Their measured cross section of 42,000 barns is
significantly smaller than the Anholt and Becker calculation (which includes
screening) of about 64,000 barns.  The result of the present Pb + Pb
calculation, which does not include screening, (about 58,000 barns) were
privately communicated to Krause et al. and they commented in their paper,
``With screening included\cite{ab} and scaled to a Au target, the Baltz value
agrees with the $\sigma_i$ measured in the ionization experiment
($4.2 \times 10^4$ b).'' 

\section{THE TWO CENTER LIGHT CONE CALCULATION OF CONTINUUM PAIRS}
Because the singular gauge involves $\delta$ functions in $(z - t)$ and
$(z + t)$, the amplitude for electron positron pair production can be
evaluated in closed form\cite{bm}.  The technique used builds on the solution
of Eq.(2) and involves conventional Green's functions methods.  The amplitude
for electron positron pair production can also be evaluated in closed form
in the light cone gauge using boundary condition considerations.  The
amplitudes obtained in both gauges agree.

With a reasonable ansatz for the physical infrared cutoff (i.e. the
factors of $\omega^2 / \gamma^2$), the closed form
result for the amplitude takes the following form 
\begin{eqnarray}
M(p,q)&=& 4 \eta^2
\int d^2 k_{\perp} \ e^{ i b k_{\perp}}\  
\biggl(\Bigl( [p_{\perp} - k_{\perp}]^2 + {\omega^2 \over \gamma^2}\Bigr)
\Bigl( [k_{\perp} + q_{\perp}]^2 + {\omega^2 \over \gamma^2}\Bigr)
\biggr)^{i \eta - 1}\nonumber \\
& \times & \biggl({ \bar{u}(p,s_f) (1-\alpha_z) ( - \not\!k_{\perp}
 + m)  v(q,s_i) \over  2 p^+ q^- + k_{\perp}^2 + m^2}
\nonumber \\
&& +\  { \bar{u}(p,s_f) (1+\alpha_z)  ( - \not\!p_{\perp} +
 \not\!q_{\perp} + \not\!k_{\perp} + m) 
 v(q,s_i) \over  2 p^- q^+ +
 (p_{\perp} - q_{\perp} -k_{\perp})^2 + m^2} \biggr). 
\end{eqnarray}
The cutoff comes in response to the spatial region 
$\mbox{\boldmath $ \rho$}   = \gamma / \omega$ where both the singular and
light cone potentials begin to lose their validity.  Without the $i \eta$ in
the exponent Eq.(3) goes over into the corresponding perturbation theory result
of Bottcher and Strayer\cite{stra}.  In fact, if one squares the exact
amplitude Eq.(3) and performs the impact parameter integral before the
$k_{\perp}$ integral, the perturbation theory result is obtained.
Thus exact cross sections to specific final states are the same as
those of perturbation theory.

On the other hand, if we define an impact parameter dependent probability
variable 
\begin{equation}
P(b) = \sum_{p,q,spins} \rho \ \vert M(p,q) \vert^2
\end{equation}
with $\rho$ the density of states, then $P(b)$ can be understood as the mean
number in a Poisson
distribution for the number of pairs N created at impact parameter $b$
\cite{em,rbw,baur}
\begin{equation}
P(N,b) = {[P(b)]^N \over N!} e^{-P(b)}
\end{equation}
The $k_{\perp}$ integral must first be performed if $P(b)$ is to be used in
Eq.(5), and one
finds that multiplicity rates should be reduced from perturbation theory.

In recent parallel work Segev and Wells\cite{sw} also note that the exact
continuum pair cross section is identical to that of perturbation theory.
They go on to note that CERN SPS date of Vane et al.\cite{vane} are consistent
with perturbation theory: specifically that the cross section for continuum
pairs goes as ${\rm Z_P^2 Z_T^2}$.  Correspondingly, Hencken, Trautmann, and
Baur\cite{he} calculate
multiple pair cross sections reduced from perturbation theory. 

\section{MUTUAL COULOMB DISSOCIATION}
The proposed zero degree calorimeter at RHIC will detect the total
neutral energy in very constrained cones downstream of beam crossing.
For neutrons of the beam momentum, the detected
signal then varies as the number of neutrons.

Detection of neutrons from
the correlated forward-backward Coulomb or nuclear dissociation of two
colliding nuclei provides a practical luminoity monitor,
independent of individual detector set-up, in heavy ion
colliders.  Single or mutual Coulomb dissociation can be reliably predicted
in a Weizsacker-Williams formalism using measured photo-dissociation data
as input.  Nuclear dissociation is geometric.  

We have calculated\cite{bcw} the total
nuclear plus Coulomb correlated dissociation cross section to be 11.0 barns
(in comparison to the 7 barn pure hadronic dissociation cross section) for
Au + Au at RHIC.  The corresponding cross section for detecting a single
neutron both forward and backward at the beam momentum is 0.45 barns.  These
single neutrons come from Coulomb excitation of the giant dipole resonance and
in coincidence provide a clean, well predicted signal useful as a beam
luminosity monitor.  

Similar slightly larger corresponding cross sections are
predicted for Pb + Pb at LHC: 0.53 barns for forward-backward single neutrons
and 14.8 barns for total nuclear plus Coulomb correlated dissociation.

Other predicted cross section for RHIC (LHC) have been calculated:
dissociation of one of the ions, 102 (227) barns; dissociation of an ion
going to
one neutron, 49 (106) barns; one neutron in one detector and any neutral
in the other 1.35 (1.88) barns.
These calculated cross sections could provide complementary roles in luminosity
monitoring, especially by exploiting their predicted ratios.

\section{SUMMARY AND CONCLUSIONS}
Distinctive physical consequences have been shown to arise from the exact
ultrareltivistic, semi-classical solution of the Dirac Equation in the
$\delta (z -t)$ gauge:

(1) There is no non-perturbative enhancement of bound-electron positron pair
production in ultrarelativistic heavy ion collisions.

(2) The beam energy dependent ($\sim \ln \gamma$) part of the ionization
cross section decreases with increasing Z of the nucleus ionized.

(3) Exact calculation of continuum electron-positron pair production is equal
to the perturbation theory result.

(4) Electron positron multiple pair production is reduced from the perturbation
theory result.

Mutual Coulomb dissociation has been suggested as a practical luminosity
monitor for RHIC and LHC.

\ 

This manuscript has been authored under Contract No. DE-AC02-98CH10886 with
the U. S. Department of Energy.

\end{document}